# Room Temperature Ferromagnetic Semiconductor Rutile Ti$_{1-x}$Co$_x$O$_{2-\delta}$ Epitaxial Thin Films Grown by Sputtering Method

Takashi Yamasaki,[1] Tomoteru Fukumura,[1] Masaki Nakano,[1] Kazunori Ueno,[2] Masashi Kawasaki[1,2,3]

[1] *Institute for Materials Research, Tohoku University, Sendai 980-8577, Japan.*

[2] *WPI Advanced Institute for Materials Research, Tohoku University, Sendai 980-8577, Japan.*

[3] *CREST, Japan Science and Technology Agency, Tokyo 102-0075, Japan.*

Room temperature ferromagnetic semiconductor rutile Ti$_{1-x}$Co$_x$O$_{2-\delta}$ (101) epitaxial thin films were grown on *r*-sapphire substrates by a dc sputtering method. Ferromagnetic magnetization, magnetic circular dichroism, and anomalous Hall effect were clearly observed at room temperature in sputter-grown films for the first time. The magnetization value is nearly as large as 3$\mu_B$/Co that is consistent with the high spin state Co$^{2+}$ in this compound recently established by spectroscopic methods. Consequently, its originally large magneto-optical response is further enhanced.

A room temperature (RT) ferromagnetic semiconductor $Ti_{1-x}Co_xO_{2-\delta}$ is one of the promising materials for semiconductor spintronics,[1,2] exhibiting large magneto-optical effect[3,4] and anomalous Hall effect[5-9] with high Curie temperature ~600 K.[10] Its device implementation has also been demonstrated as a tunneling magnetoresistance device operable at up to 200 K,[11] that is the highest operation temperature among spintronics devices using ferromagnetic semiconductors. Until now, several sophisticated methods such as pulsed laser deposition method (PLD) and molecular beam epitaxy have been exploited to synthesize $Ti_{1-x}Co_xO_{2-\delta}$ since the resultant high quality epitaxial thin films are useful to reveal its unresolved fundamental properties as a ferromagnetic semiconductor. In order to make industrial applications more plausible, however, prevailing methods such as sputtering deposition are preferred. For example, various materials can be deposited in large area uniformly by using the sputtering method for fabrication of transparent conducting oxide films and magnetic-multilayer devices. The sputtering methods have been tried to grow $Ti_{1-x}Co_xO_{2-\delta}$ films so far, but only magnetization measurement was reported without examining presence of ferromagnetically spin polarized carriers via magneto-optical and anomalous Hall effects and so on.[12-18] Some of those films were subject to the magnetic precipitations or the lack of electrical conductivity, hence it is difficult to examine fundamental properties.

In this study, rutile $Ti_{1-x}Co_xO_{2-\delta}$ epitaxial thin films were grown by using a sputtering method. Ferromagnetic magnetization, magneto-optical effect, and anomalous Hall effect were observed at room temperature in sputter-grown films for the first time. In addition, the saturation magnetization is the maximum among ever reported values and is close to the ideal value ~$3\mu_B$/Co assuming recently established high spin $Co^{2+}$ state in this compound.

Rutile $Ti_{1-x}Co_xO_{2-\delta}$ epitaxial thin films were grown on *r*-sapphire single-crystal substrates by dc magnetron sputtering method. Sputtering targets were fabricated from mixture of $TiO_2$ and $CoO$ powders to form $Ti_{1-x}Co_xO_2$ ceramic disks of 2 inch$\phi$ (4N in purity) with different *x* (*x* = 0.00, 0.05, 0.10). The target to substrate distance was 15 cm. The base pressure of the deposition chamber was $3 \times 10^{-8}$ Torr. During the growth, Ar (6N) gas was supplied at a rate of 340 standard-$cm^3$/minutes and the total pressure was $5 \times 10^{-3}$ Torr. The films were grown at 400 ˚C with sputtering dc power of 100 W. The thickness of the films was 80-110 nm. The crystal structure and Co content were determined, respectively, by powder x-ray diffraction and electron probe microanalysis. The surface morphology was examined using atomic force microscope and scanning electron microscope. The magnetization was measured in magnetic field normal to the film plane with superconducting quantum interference device magnetometer. The transmission and reflectance spectra were measured by a standard spectrometer to deduce the absorption spectra of the films. Magnetic circular dichroism (MCD) was measured with a magnetic field normal to the film plane (Faraday configuration). The resistivity and Hall effect were measured for photolithographically patterned Hall bars (60 μm wide × 220 μm long). Electron density ($n_e$) was deduced from normal part of the Hall resistance, i.e., the linear slope component with respect to the magnetic field.

Figure 1 shows x-ray diffraction patterns for $Ti_{1-x}Co_xO_{2-\delta}$ (*x* = 0, 0.05, 0.10) thin films. The orientation of the films is (101) with rutile structure without any impurity phases.[19] The lattice constants along (101) of all samples are about 2.49 Å that is close to the value of the bulk $TiO_2$ (2.487 Å) and the PLD-grown films previously reported[5] as shown in inset of Fig. 1(a). The scanning electron microscopy and atomic force microscopy confirmed no surface precipitations. The root mean square roughness of the films was within several nanometers in area of 5 μm square. From the electron probe microanalysis, the Co content was revealed to be the same as that of prescribed Co content in the targets.

The growth in Ar gas yields in electrically conducting films. Figure 2 shows electric properties for $Ti_{1-x}Co_xO_{2-\delta}$ films. The temperature dependence of resistivity ($\rho_{xx}$) shows semiconducting behavior. The mobility ($\mu$) shows small temperature dependence, while the $n_e$ shows freezing-out behavior with decreasing temperature. These values are comparable with those of PLD-grown films.[20] For $x = 0.10$, the mobility decreases a little probably due to degraded crystalline quality and/or increased impurity scattering.

Magnetization curves of the $Ti_{1-x}Co_xO_{2-\delta}$ films at 300 K are shown in Fig. 3(a). The saturation magnetization is close to 3 $\mu_B$/Co and is almost temperature independent below 300 K as shown in inset of Fig. 3(a), representing much higher Curie temperature than room temperature. It is noted that the magnetization value is evidently larger than that of Co metal (1.7 $\mu_B$/Co) ruling out possible precipitation of the Co metal. As depicted in Fig. 3(b), $x = 0.05$ and 0.10 films show large MCD amplitude in visible-ultraviolet region, where the visible and ultraviolet regions have opposite signs in MCD response with the similar amplitude.[21] The spectral feature is similar to that of PLD-grown films, while the amplitude is significantly enhanced by a factor of two.[4] From Fig. 3(c), visible absorption is negligible for $x = 0$, whereas higher $x$ results in emergence of in-gap absorption mainly due to *d-d* transition of Co ions contributing to this large MCD signal.[22]

The observed saturation magnetization is nearly twice as large as that of high quality PLD-grown (101) epitaxial films on *r*-sapphire substrates (Fig. 3(d)).[23] The saturation magnetization is close to an ideal value assuming $Co^{2+}$ with high spin state in $Ti_{1-x}Co_xO_{2-\delta}$, as was confirmed by x-ray photoemission spectroscopy and x-ray magnetic circular dichroism measurements.[24,25] In comparison with the PLD-grown films, as a result, the MCD is enhanced to be 14000 deg/cm in maximum (Fig. 3(e)).

Figure 4(a) shows the magnetic-field dependence of Hall resistivity ($\rho_H$) at 300 K for $Ti_{1-x}Co_xO_{2-\delta}$ films. For $x = 0$, only normal Hall term proportional to magnetic field is seen, and anomalous Hall term appears with finite $x$. The slope of the normal Hall term shows a slight change with respect to $x$ representing almost the same $n_e$ in each film, whereas the increase of the anomalous Hall term with $x$ is observed due to increase in the volume magnetization. With decreasing temperature, the freezing-out of $n_e$ as shown in Fig. 2 (bottom) results in rapid increase in the normal Hall term (Figs. 4(b) and (c)). A small finite hysteresis at 100 K for $x = 0.10$ in Fig. 4(c) was also observed in its magnetization curve. To summarize, the Hall effect in these films follows well-known empirical formula of Hall effect in ferromagnetic metals, $\rho_H = R_0 \mu_0 H + R_S M$, where $R_0$ is the normal Hall coefficient inversely proportional to carrier density, $\mu_0$ is the vacuum permeability, $H$ is the magnetic field, $R_S$ is the anomalous Hall coefficient, and $M$ is the magnetization. Such systematic behavior has been already observed in the PLD-grown films and is well manifested by a scaling law between anomalous Hall conductivity $\sigma_{AH}$ and longitudinal conductivity $\sigma_{xx}$, approximately $\sigma_{AH} \propto \sigma_{xx}^{1.6}$, that was firstly observed in the PLD-grown rutile $Ti_{1-x}Co_xO_{2-\delta}$ films[5] and is commonly observed in various ferromagnetic metals and semiconductors.[26] Figure 4(d) shows this relation deduced in this study that is superposed with our data of the PLD-grown rutile and anatase films.[26] The sputter-grown films show good coincidence with those PLD-grown films implying possible manipulation of the anomalous Hall effect by control of conductivity through field effect and so on.

In conclusion, $Ti_{1-x}Co_xO_{2-\delta}$ epitaxial films grown by sputtering method show ferromagnetic magnetization, giant magneto-optical effect, and anomalous Hall effect at room temperature, representing its intrinsic nature of a ferromagnetic semiconductor. The magnetization nearly as large as $Co^{2+}$ with high spin state suggests that sputter-grown films have robust ferromagnetic properties, and might be related with a local lattice distortion in the sputter-grown films recently suggested as a possible origin of the high spin state.[20]

We acknowledge F. Matsukura and H. Ohno for magnetic circular dichroism measurements. This work was partly supported by the New Energy and Industrial Technology Development Organization, the Industrial Technology Research Grant Program (05A24020d), Grant-in-Aid for Scientific Research on Priority Areas (16076205) and for Young Scientists (A19686021).

**Figure captions**

Fig. 1. X-ray diffraction patterns for $Ti_{1-x}Co_xO_{2-\delta}$ films ($x$ = 0, 0.05, 0.10) on $r$-sapphire substrates. The peaks from the substrate were skipped during the measurements. R and S denote rutile phase and substrate, respectively. Inset shows $x$ dependence of lattice constant along (101) direction ($d$ (101)) (solid circles). The data of the PLD-grown films (open circles)[5] with different growth oxygen pressures ($P_{O2}$) and the bulk specimen (open square) are also plotted for a comparison.

Fig. 2. Temperature dependence of resistivity ($\rho_{xx}$), mobility ($\mu$) and electron density ($n_e$) for $Ti_{1-x}Co_xO_{2-\delta}$ films ($x$ = 0. 0.05, 0.10).

Fig. 3. (a) Magnetic field dependence of magnetization ($M$) for $Ti_{1-x}Co_xO_{2-\delta}$ films ($x$ = 0.05, 0.10) at 300 K. Inset shows temperature dependence of the saturation magnetization ($M_S$). (b) MCD at a magnetic field of 1 T with Faraday configuration and (c) optical absorption spectra at 300 K. (d) $M_S$ and (e) maximum amplitude of MCD at 300 K as a function of $x$ (solid circles). Data of PLD-grown films are plotted for a comparison (open circles). [23]

Fig. 4. (a) Magnetic field dependence of Hall resistivity ($\rho_H$) for $Ti_{1-x}Co_xO_{2-\delta}$ films ($x$ = 0, 0.05, 0.10) at 300 K. The same curves for (b) $x$ = 0.05 and (c) $x$ = 0.10 at different temperatures. (d) The relation between the anomalous Hall conductivity ($\sigma_{AH}$) and longitudinal conductivity ($\sigma_{xx}$) for $x$ = 0.05 and 0.10 deduced from the data in (b) and (c) (solid circles). Data for PLD-grown rutile (shaded circles) and anatase (shaded

squares) $Ti_{1-x}Co_xO_{2-\delta}$ films are also plotted.[26]

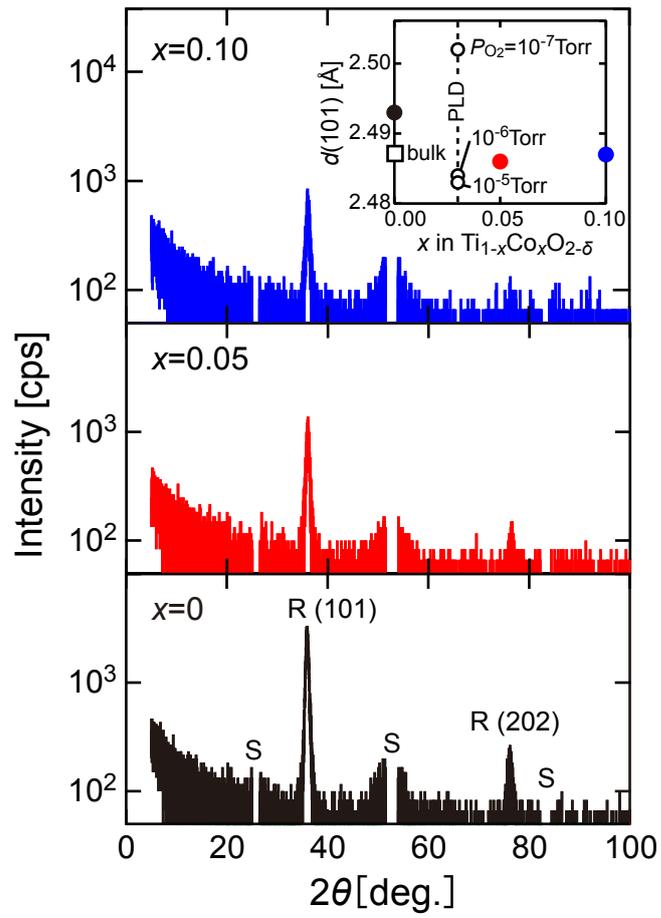

Figure 1

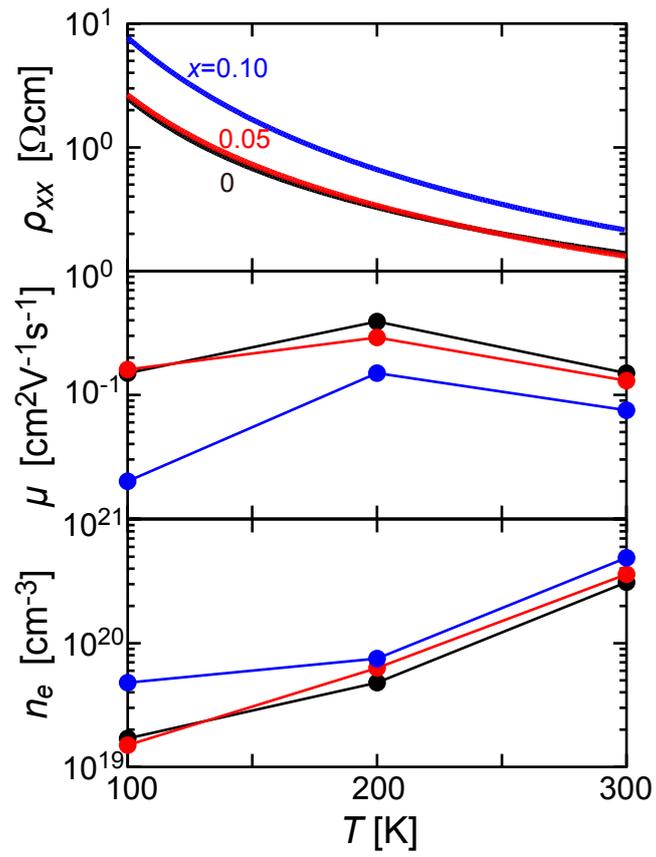

Figure 2

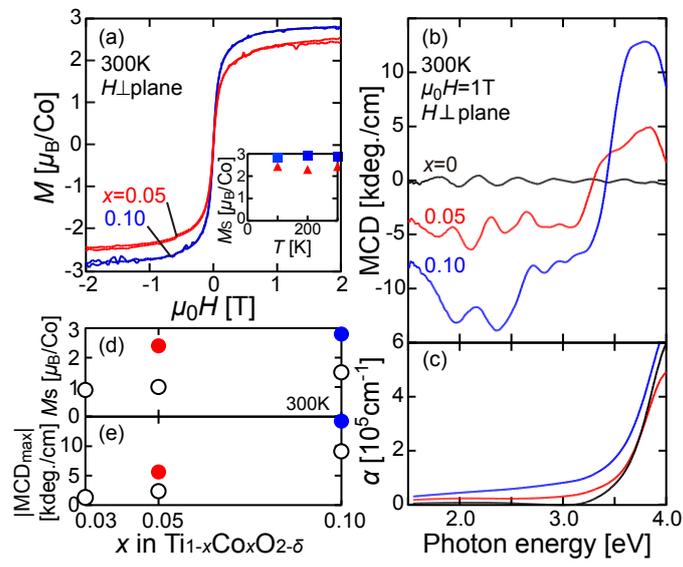

Figure 3

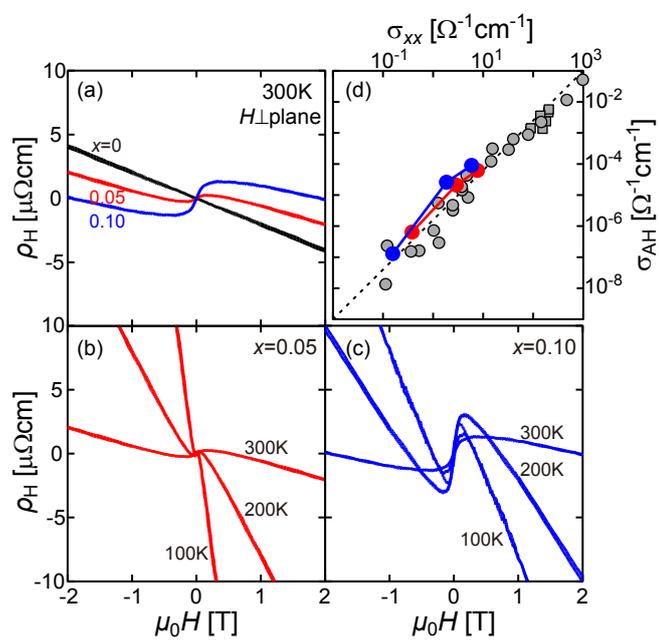

Figure 4